\documentclass{article}

\usepackage{amsmath,amsfonts,amsthm,amssymb}
\usepackage{geometry}

\usepackage{graphicx, xcolor} 
\usepackage{subfigure} 

\usepackage{caption}

\usepackage{MnSymbol}
\usepackage{stmaryrd}
\usepackage{wrapfig}
\usepackage{multirow}
\usepackage[mathscr]{euscript}
\usepackage[shortlabels]{enumitem}
\setlist{nolistsep}
\usepackage{array}
\usepackage{diagbox}
\usepackage{float}
\usepackage{framed}

\usepackage[utf8]{inputenc} 
\usepackage[T1]{fontenc}    
\usepackage{hyperref}       
\usepackage{url}            
\usepackage{booktabs}       
\usepackage{amsfonts}       
\usepackage{nicefrac}       
\usepackage{microtype}      

\usepackage{verbatim}
\usepackage{algorithm}
\usepackage{algorithmic}
\usepackage{cleveref}

\title{Laws of thermodynamics for exponential families}

\author{%
  Akshay Balsubramani \\
  \texttt{akshay@vac.bio} \\
}

\begin{document}

\maketitle

\begin{abstract}
We develop the laws of thermodynamics in terms of general exponential families. 
By casting learning (log-loss minimization) problems in max-entropy and statistical mechanics terms, we translate thermodynamics results to learning scenarios. 
We extend the well-known way in which exponential families characterize thermodynamic and learning equilibria. 
Basic ideas of work and heat, and advanced concepts of thermodynamic cycles and equipartition of energy, find exact and useful counterparts in AI / statistics terms. 
These ideas have broad implications for quantifying and addressing distribution shift. 
\end{abstract}


\section{Introduction}

Most learning problems can be solved by minimization of log loss. 
This bare fact is inescapable in modern AI and machine learning -- the variety is in the details. 
What is the space of measured data? What is the support of the distribution? 
Changing such properties of the problem fundamentally changes learning behavior, leading to the variety of modeling approaches successfully used in data science. 
But for many inference and decision-making tasks, log loss can be axiomatically inescapable. 

We explore such loss minimization problems in the language of statistical mechanics, which studies how systems of "particles" like atoms can be approximately described by relatively few bulk properties. 
There is a direct analogue to modeling, where large datasets are described by relatively few model parameters. 
This direct correspondence -- between AI-relevant loss minimization and statistical mechanics -- promises to be a rich source of insights into modeling, as it has been in various ways in the past \cite{maurer2012thermodynamics, bahri2020statistical, lamont2019correspondence, balsubramani2024entropy}. 

We take a fresh perspective on this problem, following a prescription from decades ago \cite{khinchin1949mathematical}, that the laws of thermodynamics can be derived purely from statistical-mechanical considerations: 

\begin{quote}
"The science of thermodynamics is based essentially on its two fundamental laws; thus, \textbf{every theory pretending to represent the foundation of thermodynamics must prove that these two fundamental laws can be derived from its basic principles}. Once this is done, the entire system of thermodynamic theory can be developed logically as a consequence of the two laws."
\end{quote}

Following along these lines, we develop the formal equivalents of the laws of thermodynamics in probabilistic modeling settings. 
When appropriately instantiated for physical systems, they recover the commonly known laws of thermodynamics for those systems. 

The general forms of these laws involve exponential family distributions, which link our modeling scenarios with prior physics work. 
Deriving the laws of thermodynamics for general modeling situations allows us to broaden long-developed physical intuitions to the modeling world. 

As we show, the physical counterpart of log loss in modeling is energy, and a model is described by observations on a dataset. 
Therefore, the thermodynamics of systems describe changes in these quantities, and directly govern issues of domain shift, generalization, and robustness. 
Intuitions about work and heat can be thought of in terms of changes in loss, in a completely general and rigorous manner. 
This opens the door for a classical arsenal of thermodynamic tools to be applied in machine learning.

\section{Perspectives from several fields}

This is a topic that has been at the crossroads of several fields -- physics (statistical mechanics), statistics, information theory, and more.

\subsection{Statistical physics}

These ideas first developed in physics, and their application to physics is often called "statistical physics" or "statistical mechanics". 
The main motivations for this have been the multi-particle nature of matter and bulk observations of its state, along with evident observed postulates like continuity and conservation rules. 
These have given rise to thermodynamics, with incredible consequences beginning with the refinement of engines for transport in the 1800s. 

It was quickly seen (since \cite{maxwell1904theory, gibbs1902elementary}) that statistical mechanics is a formal tool with few assumptions, and therefore applies extremely generally within and beyond physics. 
But statistical mechanics has proceeded in a unique way, from observations and postulates that are obvious in our common physical world. 
Its central concepts, of heat and energy, are intuitively understood from our experience, but remain difficult to tangibly define. 
Distributions and datasets can be very different from bulk matter systems, encompassing a much broader range of scenarios, sometimes defying apparent intuition and motivation. 
Extending statistical mechanics techniques to these scenarios has been the focus of other, more recent fields, from statistics and information theory onwards. 

Around the time these fields were forming, the same topic was studied by interdisciplinary scientists better known as leading physicists and mathematicians (e.g. \cite{schrodinger1948statistical, fisher1922mathematical, khinchin1949mathematical}). 
All found probabilistic formulations to be the most general, but did not go so far as to develop the thermodynamic laws in them. 

The key results of statistical physics concern exponential family ("Gibbs") distributions and their implications \cite{landaulifshitz}, so exponential families constitute a deep link between modeling and statistical mechanics \cite{grunwald2007minimum}. 
A crown jewel of statistical physics is classical thermodynamics -- the study of the properties of macroscopic systems (datasets) in statistical mechanical equilibrium. 

More recently, many amazing discoveries have been made in non-equilibrium thermodynamics \cite{crooks1999entropy, jarzynski1997nonequilibrium}. 
Though the terminology and focus of statistical modeling scenarios have diverged significantly now from physics, many of those physics discoveries are again implicit statements about exponential families, and they can be interpreted in terms of the laws we derive.

\subsection{Statistics}

Statistics and machine learning have not developed versions of the laws, but there has been widespread successful use of exponential families. 
These have been thought of not only as distributions that maximize entropy under observed constraints, but also in terms of sufficiency and robustness in light of those observations. 

Exponential families were discovered in statistics, as exactly the distributions with sufficient statistics whose dimension does not grow with the amount of data (i.e., they have a fixed number of parameters rather than being a nonparametric model). 
This foundational result is known as the Pitman-Koopman-Darmois theorem on sufficient statistics \cite{fisher1922mathematical, koopman1936distributions, Darmois1935, pitman1936sufficient}. 
A similarly central place is taken by the factorization theorem for probability densities characterizing sufficient statistics, also studied in depth around that time \cite{neyman1935suunteorema} and shown to be a very general perspective on sufficiency \cite{halmos1949application} 
\footnote{
In this case, the factorization theorem \cite{besag1974spatial} says that any statistic $T(X)$ is sufficient if and only if the probability mass function (PMF) of $X$ can be factored as $f_n (X|\theta) = u(X)v(T(X), \theta)$ for some functions $u$ and $v$. 


}, 
shown to hold for exponential families. 


The favorable properties of exponential families in terms of conditioning, Bayesian inference and conjugacy, and statistical efficiency have made them central characters in statistics \cite{wainwright2008graphical}. 
More modern perspectives on exponential families have developed more game-theoretic robustness characterizations \cite{GD04}, as discussed in the next section.

\subsection{Theories of information, games, and optimization}

Drawing from the foundational work in statistical mechanics, there are deep connections between loss minimization and information-theoretic quantities. 
The link begins with Boltzmann's original "probability calculations," which showed how the key quantities of information theory, including entropy, emerge naturally when computing the likelihood of observing particular distributions of samples. 
These same calculations reveal that high-entropy configurations are exponentially more likely than other configurations, providing a rigorous grounding for maximum entropy principles in learning. 
Information theory work has generalized Boltzmann's probability calculations \cite{csiszar1984sanov}, with deep impacts in probability theory \cite{dembozeitouni2009large}. 
Exponential families are found to be fundamental to information theory, with efficient practical structures like trees and  sparse connectivity, admitting practically informative factorizations and efficient algorithms in fundamental ways \cite{chow1968approximating}. 

Information geometry \cite{amari2001information} is a noteworthy synthesis of many of these ideas in very general probabilistic settings. 
A large quantity of related work brings information theory and statistics together, with exponential families and statistical mechanics techniques playing a key role \cite{efron1978geometry}. 
Physicists since Jaynes \cite{jaynes1957information, jaynes1957information2, jaynes1986monkeys} attempted to carry these hard-won physics results and intuitions forward into information theory. 
In parallel, we have seen revived interest in rigorously extending fully-formed principles of statistical physics to information theory \cite{csiszar2004information, DPDPL97}. 
The laws of thermodynamics have not been completely formulated in such settings, however.

Exponential families have been further equipped with a game-theoretic motivation in more recent times \cite{topsoe1979information, grunwald2007minimum}. 
This perspective, closely related to convex optimization, motivates exponential families as being the most robust for prediction models under log loss. 
Though all the ways of looking at exponential families are equally true and relevant, this one fundamentally underpins how we derive the thermodynamic laws.

\subsection{Outline of this paper}

This paper works within the statistical mechanics perspective on modeling with loss minimization, where it corresponds to modeling with exponential families. 
The resulting AI/statistics theory that we develop follows classical thermodynamics closely. 
While the physics is driven by carefully chosen bulk observations (feature functions), the exponential family perspective is typically much more general. 
The probabilistic statements we develop apply in AI / statistics settings that extend the corresponding physics statements in interesting and sometimes unexpected ways. 
Many principles of statistical thermodynamics were unified in a final form before the advent of information theory. 
But information theory revived interest in rigorously connecting purely probabilistic notions of entropy to the original statistical physics version. 
This link was stressed by Jaynes in vocally advocating maximum entropy methods for statistical inference \cite{jaynes1957information, jaynes1957information2}. 
Much has changed since then, with statistically powered methods flourishing in large-data scenarios. 

The results in this work are a link from statistical mechanics to modern AI, deepening the original connections of Jaynes' earlier work to information theory. 
We find that just as exponential families are good generic distributions for learning and correspond to statistical mechanics equilibria, their performance under model misspecification and distribution shift corresponds to the study of thermodynamics. 
Here we begin developing this subject, exploring what these ideas mean as a set of powerful and principled tools for understanding models, and laying the groundwork for further such investigation. 

\Cref{sec:expfamprimer} defines some common statistical ground between information theory and statistical physics, in physics terms. 
\Cref{sec:lawswarmup} and \Cref{sec:lawsofthermo} build upon this framework, deriving analogues of the laws of thermodynamics in it: results that concern how modeling loss changes when an exponential family equilibrium is shifted.
\Cref{sec:fluctuations} develops more related thermodynamics concepts in probabilistic terms, concerning fluctuations around an exponential family equilibrium. 
These are all discussed and anchored in more physics intuitions in \Cref{sec:discussion}.

\section{Exponential families with statistical mechanics: a primer}
\label{sec:expfamprimer}

We summarize exponential families and some of their properties from a rigorous statistical mechanics viewpoint, identifying a useful framework of connections between statistical physics, information theory, and learning \cite{balsubramani2024entropy}. 

This centers around the quantities $\text{H} (P), \text{H} (P, Q), \text{D} (P \Vert Q)$, traditionally thought of as being from information theory. 
$\text{H} (P, Q)$ is the log loss of $Q$ to $P$ -- the quantity being minimized in learning problems. 
Meanwhile, the quantities 
\begin{align*}
\text{H} (P) = \min_{S \in \Delta (\mathcal{X})} \text{H} (P, S) 
\qquad
\text{D} (P \Vert Q) = \text{H} (P, Q) - \text{H} (P)
\end{align*} 
are nicely interpretable. 
$\text{H} (P)$ is the Bayes loss -- the minimum loss suffered by any predictor on this problem because of the indeterminacy in the data. 
The divergence $\text{D} (P \Vert Q)$ is evidently interpretable as the regret with respect to the inherent (Bayes) loss in describing the data $P$.

\subsection{Basic setup}

Exponential families project onto a set of feature moment constraints. 
\begin{itemize}
\item 
We know the data space $\mathcal{X}$, and we have a set of feature functions $f_i : \mathcal{X} \to \mathbb{R}$, $i = 1, \ldots, d$ that we can observe over $\mathcal{X}$. 

\item 
We sample $n$ elements from $\mathcal{X}$ using an unknown distribution $P$, giving an empirical measure $\hat{P}_{n}$. 

\item
We observe the expected values of these features $f_i$ over some data distribution $\hat{P}_{n}$, $\mathbb{E}_{x \sim \hat{P}_{n}} [f_i (x)] = \alpha_i$, $i = 1, \ldots, d$. 
So $\hat{P}_{n} \in \mathcal{A}$, where again remember that 
\begin{align*}
\mathcal{A} := \left\{ P \in \Delta (\mathcal{X}): \mathbb{E}_{x \sim P} [ f_i(x) ] = \alpha_i \;\forall i = 1, \dots, d \right\}
\end{align*}
\end{itemize}

The information projection of $P$ on $\mathcal{A}$, $P_{\mathcal{A}}^{*}$, is the closest distribution to $P$ in $\mathcal{A}$ according to $\text{D} (\cdot \Vert \cdot)$, which has an exponential family form:  
\begin{align*}
\boxed{
P_{\mathcal{A}}^{*} := \arg\min_{Q \in \mathcal{A}} \text{D} (Q \Vert P) 
\iff 
P_{\mathcal{A}}^{*} (x) \propto P (x) \exp \left( \sum_{i=1}^{d} \lambda_i^* f_i(x) \right)
}
\end{align*} 

where $\left\{ \lambda_i^* \right\}_{i=1}^{d}$ are the Lagrange multipliers that ensure $P_{\mathcal{A}}^{*} \in \mathcal{A}$.\footnote{We'll assume these exist, so that the problem is well-posed. (See \cite{follmer2011stochastic}, Thm. 3.24 for the calculation.)} 

The normalization is expressed through the \textit{log-partition function}
$ \displaystyle
\text{A} (\lambda) = \log \mathbb{E}_{P} \left[ \exp \left( \sum_{i=1}^{d} \lambda_i f_i(x) \right) \right]
$, 
so that $P_{\mathcal{A}}^{*} (x) = P (x) \exp \left( \sum_{i=1}^{d} \lambda_i^* f_i(x) - \text{A} (\lambda^{*}) \right)$.

$P_{\mathcal{A}}^{*}$ is unique in all non-degenerate cases, and has a convenient closed form parametrized by $\lambda \in \mathbb{R}^{d}$. 
The family of such distributions with varying $\lambda$ is called the exponential family associated to features $f (x) \in \mathbb{R}^{d}$ and prior $P$:\footnote{If no prior is given, a uniform prior is typically assumed.} 
\begin{align*}
\mathcal{Q}
:= 
\left\{ Q (x \mid \lambda) \in \Delta(\mathcal{X}): \exists \lambda \in \mathbb{R}^{d} : \;
Q (x \mid \lambda) \propto P (x) \exp \left( \sum_{i=1}^{d} \lambda_{i} f_{i} (x) \right) \right\}
\end{align*}

From these definitions, observe that $P_{\mathcal{A}}^{*} \in \mathcal{A} \cap \mathcal{Q}$ \footnote{We'll ignore issues of $\mathcal{Q}$ being an open/closed set here; otherwise all the results we show are true with $\mathcal{Q}$ being replaced by its closure.} -- the information projection is in the exponential family, and follows the observed constraints $\mathcal{A}$.

\subsection{Internal energy and the log-partition function}

Define the \textbf{internal energy} that the exponential family associates with any microstate $x$ and feature $i$ as 
$$ U_i^{\lambda} (x) := - \lambda_i f_i (x)$$ 
This is the energy of the system in state $x$ associated with the $i$-th feature, which adds to make the total energy $ \text{U}^{\lambda} (x) := \sum_{i=1}^{d} U_i^{\lambda} (x) $, 
so that $P_{\lambda} (x) \propto \exp \left( - \sum_{i=1}^{d} U_i^{\lambda} (x) \right) = \exp \left( - \text{U}^{\lambda} (x) \right)$. 
Taken over distributions, we have 
$$ \text{U}^{\lambda} (P) := - \sum_{i=1}^{d} \lambda_i \mathbb{E}_{x \sim P} [f_i (x)] = \mathbb{E}_{x \sim P} \left[ \sum_{i=1}^{d} U_i^{\lambda} (x) \right] $$ 

At equilibrium (when the expected moments match the empirical moments), the internal energy is $- \sum_{i=1}^{d} \lambda_i \alpha_i$ where $\alpha_i$ are the supplied constraints. 

The log-partition function is:
\begin{align}
\label{eq:logpartitionminimax}
\text{A} ( \lambda ) &:= \min_{Q \in \Delta(\mathcal{X})} \;\max_{P \in \Delta(\mathcal{X})} \left[ \text{H} (P, Q) - \text{U}^{\lambda} (P) \right]
\end{align}
(We have seen that this means $\displaystyle \text{A} (\lambda) = \log \mathbb{E}_{P} \left[ \exp \left( \sum_{i=1}^{d} \lambda_i f_i(x) \right) \right] $.)

This is a good way to \textit{define} the log-partition function from a machine learning perspective. 
It illuminates a central zero-sum game being played between the min-player and the max-player. 

If we use the minimax theorem \cite{S58} to swap the order of the min and max, 
\begin{align*}
\text{A} ( \lambda ) &= \max_{P \in \Delta(\mathcal{X})} \;\min_{Q \in \Delta(\mathcal{X})} \left[ \text{H} (P, Q) - \text{U}^{\lambda} (P) \right] \\
&= \max_{P \in \Delta(\mathcal{X})} \left[ \text{H} (P) + \sum_{i=1}^{d} \lambda_{i} \mathbb{E}_{x \sim P} [ f_{i} (x) ] \right]
\end{align*}

Since $A$ is a maximum over linear functions of $\lambda$, it is convex in $\lambda$.

\subsection{Free energy}

The \textbf{free energy} of any distribution $P$ is the difference between its internal energy and its entropy: 
\begin{align*}
\text{F}^{\lambda} (P) := \text{U}^{\lambda} (P) - \text{H}(P) = - \sum_i \lambda_i \mathbb{E}_{x \sim P} [f_i (x)] - \text{H}(P)
\end{align*}

Therefore, for any exponential family distribution $P_{\lambda}$, 
\begin{align*}
\text{F}^{\lambda} (P_{\lambda}) = - \text{A} (\lambda)
\end{align*}

In the language of constrained optimization, the free energy is the Lagrangian of the constrained maximization of entropy -- indeed, $\text{A} (\lambda)$ is the Fenchel dual \cite{boyd2004convex} of the negative entropy function. 
It has a more physically intuitive interpretation too in the study of thermodynamics, involving concepts of work and heat, which will be connected to our statistical view in the next sections.

\subsection{Entropy}

In general, the entropy is a minimum of linear functions by definition -- $\text{H} (P) = \min_{S \in \Delta (\mathcal{X})} \text{H} (P, S)$ where $\text{H} (P, S) = \mathbb{E}_{x \sim P} \text{H} (x, S)$. 
Therefore, $\text{H} (P)$ is concave in $P$. 

The entropy of the exponential family distribution $P_{\lambda}$ is 
\begin{align*}
\text{H} (P_{\lambda}) 
&:= \text{U}^{\lambda} (P_{\lambda}) - \text{F}^{\lambda} (P_{\lambda}) 
= \sum_{i=1}^{d} - \lambda_i \mathbb{E}_{x \sim P_{\lambda}} \left[ f_i(x) \right] + \text{A} (\lambda)
= - \sum_{i=1}^{d} \lambda_i \alpha_i + \text{A} (\lambda)
\end{align*}

This is convex in $\lambda$; when it is being maximized over a convex constraint set $\mathcal{A}$, this means that the entropy-maximizing $P_{\lambda^{*}} = P_{\mathcal{A}}^{*}$ lies at the boundary of the set $\mathcal{A}$. 


\subsection{Loss and internal energy}
\label{sec:lossinternalenergy}

For any distribution $P$ whatsoever, and any parameters $\lambda$, the loss $\mathbb{E}_{x \sim P} \left[ - \log (P_{\lambda} (x)) \right] = \text{H} (P, P_{\lambda})$ is: 
\begin{align}
\label{eq:defofcrossent}
\text{H} (P, P_{\lambda})
=
\text{A} (\lambda) - \sum_i \lambda_i \mathbb{E}_{x \sim P} [f_i (x)]
= \text{U}^{\lambda} (P) - \text{F}^{\lambda} (P_{\lambda})
\end{align}

In other equivalent words, for any $\lambda$: 
\begin{align*}
\text{H} (x, P_{\lambda}) = - \sum_{i=1}^{d} \lambda_{i} f_{i} (x) + \text{A} (\lambda) = \text{U}^{\lambda} (x) - \text{F}^{\lambda} (P_{\lambda})
\end{align*}

(To get some intuition on this, if the loss is low, $\text{U}^{\lambda} (x) \approx \text{F}^{\lambda} (P_{\lambda})$, which is highly negative. 
So the energy is generally negative over the data.) 

Therefore, for any two arbitrary distributions $P, Q$: 
\begin{align*}
\text{H} (P, P_{\lambda}) - \text{H} (Q, P_{\lambda})
= \text{U}^{\lambda} (P) - \text{U}^{\lambda} (Q)
\end{align*}

Hence, $\text{U}^{\lambda} (P) = \text{H} (P, P_{\lambda}) + K$ for some constant $K$. \footnote{The constant must result in $U^{0} (P) = 0$, i.e. $K = - \text{H} (P, P_{0}) = - \text{H} (P_{0})$ where $P_{0} = P_{\lambda = 0}$ is the prior distribution.}

This shows that the internal energy $\text{U}^{\lambda} (P)$ of data $P$ corresponds to the loss incurred by predicting with $P_{\lambda}$ on $P$. 
It is low when $P_{\lambda}$ is a good approximation of the data, and higher otherwise.

\subsection{Regret and free energy}

Subtracting $\text{H}(P)$ from both sides of the equation \eqref{eq:defofcrossent}, we get that for any distribution $P$, 
\begin{align*}
\text{D} \left( P \Vert P_{\lambda} \right) = \text{F}^{\lambda} (P) - \text{F}^{\lambda} (P_{\lambda})
\end{align*}

For us, the free energy therefore corresponds to the model's regret. 
It is the excess (or "reducible") loss, over what would be incurred if we knew $P$. 
If the model does as well as $P$ at describing the data, the regret/free energy will be low, even if $P$ is noisy. 

In the language of duality in statistical physics, this is the dual interpretation (minimum free energy at fixed "temperature" $\lambda$) to the usual primal variational characterization (maximum entropy at fixed internal energy).

\section{Warming up to the laws}
\label{sec:lawswarmup}

We proceed as suggested in the introduction, to develop probabilistic laws of thermodynamics. 
In order to develop statistical intuition on this subject, we can recall that the cross-entropy loss ("deviance" in this context) is extremely convenient to work with under distribution shift when predicting with an exponential family model $P_{\lambda}$. 

To summarize this, suppose the data distribution's observed statistics are perturbed from $\alpha$ to $\alpha + \Delta^{\alpha}$. 
Any such change in data distribution also changes the loss, by an amount that is linear in the perturbation. 
\begin{align*}
\text{H} (\alpha + \Delta^{\alpha}, P_{\lambda}) - \text{H} (\alpha, P_{\lambda}) 
= \sum_{i=1}^{d} \lambda_{i} \Delta_{i}^{\alpha} 
\end{align*}

In this manner, the change in loss decomposes over features and over data, and contributions can be computed for each feature and each sample. 
This additive decomposition generalizes the situation for sum-of-squares loss functions, which arise when the prediction model is a standard Gaussian. 
This concept is a major advantage to using deviance, underlying Fisher's original achievement of ANOVA and other uses of generalized linear models since then \cite{MN89}. 

The laws of thermodynamics answer two questions that naturally emerge from this process: 
\begin{itemize}
\item 
How does the change in loss break down into "reducible" changes that better model training can fix, versus "irreducible" changes in the noise level of the dataset itself? 
\item 
What happens when the prediction model changes as well? 
\end{itemize}

Recalling (e.g. \Cref{sec:lossinternalenergy}) that this notion of loss corresponds to internal energy, 
the above questions can be rephrased as changes in energy -- the commonly known subject of the laws of thermodynamics.

\section{Laws of thermodynamics}
\label{sec:lawsofthermo}

In this section, we explore how the energy varies, locally and globally, when a modeling problem is changed from an exponential family equilibrium. 
The energy corresponds to prediction performance, so in modeling terms, this studies the impact of distribution shift on prediction loss.

The effect of the data on an exponential family equilibrium is driven by the data $f (x)$ and observed constraints $\alpha$. 
So suppose $f (x), \alpha$ both change. 
Each observed data point changes by a small amount $\Delta (x) := ( \Delta_{1} (x), \dots, \Delta_{d} (x)) \in \mathbb{R}^{d}$. 
The constraints change by a small amount $\Delta^{\alpha} := ( \Delta_{1}^{\alpha}, \dots, \Delta_{d}^{\alpha} ) \in \mathbb{R}^{d}$. 

In other words, 
\begin{align*}
f (x) \impliedby f (x) + \Delta (x) \qquad \alpha \impliedby \alpha + \Delta^{\alpha}
\end{align*}
In this paper, we take $\Delta (x)$ and $\Delta^{\alpha}$ to be infinitesimal, taking total differentials. 

Now we quantify how much the performance changes when modeling the distribution. 
In particular, we are interested in the performance -- the internal energy -- as well as its regret-based counterpart, the free energy. 
Recall that the free energy $\text{F}^{\lambda}$ of a distribution $P$ is the regret of predicting on $P$ with $P_{\lambda}$ -- the difference between the loss (internal energy) $\text{U}^{\lambda} (P)$ and the best achievable loss (entropy) $\text{H} (P)$: 
\begin{align}
\label{eq:freeenergygeneral}
\text{F}^{\lambda} (P) = \text{U}^{\lambda} (P) - \text{H}(P)
\end{align}

Any $\Delta (x) , \Delta^{\alpha}$ will perturb the performance, in the form of these quantities. 
We now quantify how $\Delta (x) , \Delta^{\alpha}$ affect the loss $\text{U}^{\lambda}$, 
and how this is split between the regret $\text{F}^{\lambda}$ (reducible loss) and the entropy $\text{H}$ (irreducible loss / Bayes risk).

\subsection{Thermodynamic equilibrium}

When we look at shifts that change data distributions, we must distinguish between different types of changes, which all affect prediction differently.

The core concept is thermodynamic \textbf{equilibrium}, which has an evident physical meaning. 
Its statistical meaning emerges from our discussion: given specified features $f (x)$ and constraints $\alpha$, equilibrium is a state $\lambda \in \mathbb{R}^{d}$ in which $P_{\lambda}$ solves a max-entropy problem under $f (x)$ and $\alpha$, with $\mathbb{E}_{x \sim P_{\lambda}} [f (x)] = \alpha$. 

The parameters $\lambda$ correspond to pressure, temperature, and chemical potentials. 
The features $f(x)$ correspond to measured bulk quantities like volume and particle number, and $\alpha$ to measurements of these quantities \cite{chandler1987introduction}. 

Keeping $f(x)$ constant corresponds to holding $\Delta (x) = 0$, and similarly a constant $\alpha$ means $\Delta^{\alpha} = 0$. 
Once $\alpha$ and $f (x)$ are changed, the system's performance has changed as well. 
The concepts of \textbf{heat} and \textbf{work} are central to describing this evolution in our setting, as in classical physics. 
The following sections developing the laws of thermodynamics will show what they mean in detail. 
We will see that work represents changes in regret -- "effective" performance with respect to the best achievable loss, or reducible loss. 
In contrast, heat represents changes in performance that are associated instead with the irreducible loss (which is the entropy).

\subsection{Fundamental thermodynamic relations}

We focus on the common situation in which the data can be well approximated by an exponential-family distribution $P_{\lambda}$, and characterize distribution shifts away from this situation. 
To do this, we must reexamine the defining properties of $P_{\lambda}$. 

In this situation, $P = P_{\lambda}$, so 
\begin{align}
\label{eq:freeenergyexpfam}
\text{U}^{\lambda} (P_{\lambda}) = \text{F}^{\lambda} (P_{\lambda}) + \text{H}(P_{\lambda})
\end{align}

This is a basic relationship that defines exponential families. 
It relates the free energy $\text{F}^{\lambda} (P_{\lambda}) = - \text{A} (\lambda)$, the internal energy $\text{U}^{\lambda} (P_{\lambda})$, and the entropy $\text{H} ( P_{\lambda})$. 

We proceed by writing the change in all these quantities as a function of $\Delta (x) , \Delta^{\alpha}$. 
This is the source of concepts like work and heat, as will become clear. 
The following sections will show how such concepts have immediate meaning for predictive loss-minimization modeling under distribution shift.

\subsubsection{Free energy and work}

We can start with the regret -- the performance with respect to the best achievable model. 
How much is the regret affected by $\Delta_{i} (x)$ and $\Delta_{i}^{\alpha}$ for any feature $i$? 

The regret is tantamount to the free energy, as defined earlier. 
Writing $\text{Z} := \exp (A) = \mathbb{E} \left[ \exp \left( \sum_{i=1}^{d} \lambda_i f_{i} (x) \right) \right]$, so that 
\begin{align*}
\text{d} [\text{F}^{\lambda} (P_{\lambda} )] &= - \frac{1}{\text{Z} (\lambda)} \text{d} \text{Z} (\lambda) 
= - \frac{ \sum_{i=1}^{d} \mathbb{E} \left[ \left( [\text{d} \lambda_{i}] f_{i} (x) + \lambda_{i} \Delta_{i} (x) \right) \exp \left( \sum_{i=1}^{d} \lambda_i f_i(x) \right) \right] }{\mathbb{E} \left[ \exp \left( \sum_{i=1}^{d} \lambda_i f_i(x) \right) \right]} \\
&= - \sum_{i=1}^{d} \mathbb{E}_{x \sim P_{\lambda}} \left[ \left( [\text{d} \lambda_{i}] [f_{i} (x)] + \lambda_{i} \Delta_{i} (x) \right) \right] 
= \mathbb{E}_{x \sim P_{\lambda}} \left[ \text{d} \text{U}^{\lambda} (x) \right]
\end{align*}

This is another way to understand the name of the "free energy" -- any variation in it amounts to changes in internal energy, in expectation over the equilibrium $P_{\lambda}$.

\subsubsection{Entropy and heat}

In these terms, we can also look at the change of the other term - the entropy term $\text{H} (P_{\lambda})$. 
The change in entropy -- the best achievable modeling performance on the data -- can be decomposed in terms of the changes $\Delta (x) , \Delta^{\alpha}$ . 
\begin{align*}
\text{d} \text{H} (P_{\lambda}) 
&= \text{d} [\text{A} (\lambda)] - \text{d} \left[ \sum_{i=1}^{d} \lambda_{i} \alpha_{i} \right] \\
&= \sum_{i=1}^{d} \left[ \left( [\text{d} \lambda_{i}] \mathbb{E}_{x \sim P_{\lambda}} [f_{i} (x)] + \lambda_{i} \mathbb{E}_{x \sim P_{\lambda}} \left[ \Delta_{i} (x) \right] \right) - \left( [\text{d} \lambda_{i}] \alpha_{i} + \lambda_{i} \Delta_{i}^{\alpha} \right) \right] \\
&= \sum_{i=1}^{d} \left( [\text{d} \lambda_{i}] \left( \mathbb{E}_{x \sim P_{\lambda}} [f_{i} (x)] - \alpha_{i} \right) + \lambda_{i} \left( \mathbb{E}_{x \sim P_{\lambda}} \left[ \Delta_{i} (x) \right] - \Delta_{i}^{\alpha} \right) \right) \\
&= \sum_{i=1}^{d} \left( \underbrace{[\text{d} \lambda_{i}] g^{\lambda}_{i}}_{\text{excess heat}} + \underbrace{\lambda_{i} [\delta Q_i^{\lambda}]}_{\text{housekeeping heat}} \right)
\end{align*}

where the change associated with a given feature $i$ driving the "housekeeping heat" is 
$$ \delta Q_i^{\lambda} := \mathbb{E}_{x \sim P_{\lambda}} \left[ \Delta_{i} (x) \right] - \Delta_{i}^{\alpha} $$
and the slack in the constraints for feature $i$ driving the "excess heat" is 
\begin{align*}
g^{\lambda}_{i} := \mathbb{E}_{x \sim P_{\lambda}} [f_{i} (x)] - \alpha_{i}
\end{align*}

These terms are borrowed from physics (see \Cref{sec:discussion}).

\subsection{First law}

From the above discussion, we can break down the free energy change into heat and work terms, resulting in an analogue to the first law of thermodynamics. 

Recall that 
$$ \text{U}^{\lambda} (P) := \mathbb{E}_{x \sim P} \left[ \text{U}^{\lambda} (x) \right] = - \sum_{i=1}^{d} \lambda_i \mathbb{E}_{x \sim P} [f_i (x)] $$

Therefore, any change in $\text{U}^{\lambda} (P_{\lambda})$ can be decomposed into changes in free energy (work) and entropy (heat) components. 
\begin{align*}
\text{d} \text{U}^{\lambda} (P_{\lambda}) &= \text{d} \text{F}^{\lambda} (P_{\lambda}) + \text{d} \text{H}(P_{\lambda}) \\
&= \underbrace{ \mathbb{E}_{x \sim P_{\lambda}} \left[ \text{d} \text{U}^{\lambda} (x) \right] }_{\text{Work}} + \underbrace{ \underbrace{ \sum_{i=1}^{d} \lambda_{i} [\delta Q_i^{\lambda}]}_{\text{housekeeping}} + \underbrace{ \sum_{i=1}^{d} [\text{d} \lambda_{i}] g^{\lambda}_{i} }_{\text{excess}} }_{\text{Heat}}
\end{align*}

This is the content of the first law, decomposing changes in internal energy into work and heat changes. 
(Note that $\delta Q_{i}^{\lambda}$ is not an exact differential of any quantity. )

\subsection{Second law}

The second law is perhaps more ambiguous than the first -- it describes a general pattern. 
In many interesting and practical situations, entropy tends to increase with time, and never or seldom decreases. 
The problem in physics is really to determine the conditions under which this holds \cite{baez2024entropy}. 

In our setting, the analogous problem is to formalize situations in statistics/learning under which this holds. 
There are several relevant related formalizations.

\subsubsection{Heat measures changes in entropy}

Earliest is a simple form of the second law observed in physics, by Clausius and others in the 19th century. 
As we have seen, any change in entropy can be decomposed in the following way: 
\begin{align*}
\text{d} \text{H} (P_{\lambda}) = \sum_{i=1}^{d} \left( [\text{d} \lambda_{i}] g^{\lambda}_{i} + \lambda_{i} [\delta Q_i^{\lambda}] \right)
\end{align*}
How can this be interpreted? 
If the first "excess heat" term $\sum_{i=1}^{d} [\text{d} \lambda_{i}] g^{\lambda}_{i}$ is zero, then 
\begin{align*}
\text{d} \text{H} (P_{\lambda}) = \sum_{i=1}^{d} \lambda_{i} [\delta Q_i^{\lambda}] 
\end{align*}
This happens (the excess heat term is zero) for any process in which the parameters $\lambda$ are not changing ($\text{d} \lambda = 0$). 
It also happens even if the parameters change, as long as $g^{\lambda} = 0$; such a process is called \textbf{quasistatic}. 

So for quasistatic processes, the infinitesimal entropy change is just the housekeeping heat change, and a notion of temperature for each feature $i$ in a quasistatic process is $T_{i} = 1/\lambda_{i}$. 
And when considering $d = 1$ with just a single observed feature ("energy"), the change in entropy is proportional to the heat change $\delta Q_1^{\lambda}$. 
This is the second law of thermodynamics as initially observed, in different units \cite{clausius1867mechanical, sethna2021statistical}.

\subsubsection{Work is limited by free energy}

A general microscopic statement of the second law  \cite{jarzynski1997nonequilibrium, chandler1987introduction} in our terms is: any change of the model causes a change in regret that is less than the change in performance. 
This intuitively appealing statement can be proved. 

Suppose we are given some initial conditions for the system (a training dataset), 
and the model is changed from $\lambda_{0}$ to $\lambda_{1}$, with common prior $P$ over the data. 
If we calculate the expected change in work, 
\begin{align*}
- \Delta \text{F}^{\lambda} (P_{\lambda}) 
&:= - \text{F}^{\lambda_1} (P_{\lambda_1}) + \text{F}^{\lambda_0} (P_{\lambda_0}) 
= \text{A} (\lambda_{1}) - \text{A} (\lambda_{0}) \\
&= \log \mathbb{E}_{P} \left[ \exp \left( - \text{U}^{\lambda_1} ( x ) + \text{U}^{\lambda_0} ( x ) \right) \right] \\
&= \log \mathbb{E}_{P} \left[ \exp \left( - \Delta \text{U}^{\lambda} (x) \right) \right]
\end{align*}

Rearranging, 
$$ \exp \left( - \Delta \text{F}^{\lambda} (P_{\lambda}) \right) = \mathbb{E}_{P} \left[ \exp \left( - \Delta \text{U}^{\lambda} (x) \right) \right] $$

This is known as Jarzynski's equality. 
Applying Jensen's inequality to this gives another statement of the second law: 
\begin{align}
\label{eq:secondlawjensens}
\exp \left( - \Delta \text{F}^{\lambda} (P_{\lambda}) \right) \geq \exp \left( \mathbb{E}_{P} \left[ - \Delta \text{U}^{\lambda} (x) \right] \right) 
\implies 
\boxed{\Delta \text{F}^{\lambda} (P_{\lambda}) \leq \mathbb{E}_{P} \left[ \Delta \text{U}^{\lambda} (x) \right] }
\end{align}

Of all the second law statements, this is notably general. 
The proof above holds for any model shift -- whether the models share features or not -- and any prior $P$.


\subsubsection{Interpretation: entropy production}

The first term of heat change is zero when $\mathbb{E}_{x \sim P_{\lambda}} [f (x)] = \alpha$ (quasistatic, constraints are satisfied) or when $d \lambda = 0$ (no change in parameters). 

The $\delta Q^{\lambda}$ term is zero when $\mathbb{E}_{x \sim P_{\lambda}} \left[ \Delta (x) \right] = \Delta^{\alpha}$ for all $i$ (no change in constraints beyond what is seen in the data).

For any antisymmetric feature $\omega$, we can show a similar equality. 
\begin{align*}
\mathbb{E}_{x \sim P_{\lambda}} \left[ \exp \left( - \omega \right) \right] = 1
\end{align*}

This is useful in various situations; for example, if $\omega$ is the entropy production in transitioning from $P_{I} \to P_{F}$: 
\begin{align*}
\omega := \text{H} (x, P_{F}) - \text{H} (x, P_{I}) 
\end{align*}


Now remember that $P_{i} (x) := \exp \left( \text{F}^{\lambda_{i}} (P_{\lambda_{i}}) - \text{U}^{\lambda_{i}} (x) \right)$ for $i \in \{ I, F \}$, so 
\begin{align*}
\omega &= - \text{F}^{\lambda_{F}} (P_{\lambda_{F}}) + \text{U}^{\lambda_{F}} (x) + \text{F}^{\lambda_{I}} (P_{\lambda_{I}}) - \text{U}^{\lambda_{i}} (x) 
= - \Delta \text{F}^{\lambda} (P_{\lambda}) + \Delta \text{U}^{\lambda} (x)
\end{align*}

It implies that $\mathbb{E}_{x \sim P_{\lambda}} \left[ \exp \left( - \omega \right) \right] = 1$, i.e. $\omega \geq 0$ by Jensen's inequality. 
So 
$$ \Delta \text{F}^{\lambda} (P_{\lambda}) \leq \mathbb{E}_{x \sim P_{\lambda}} \left[ \Delta \text{U}^{\lambda} (x) \right] $$

This is a different version of what we showed in \eqref{eq:secondlawjensens}. 
Versions of this argument that incorporate heat quantify entropy production in a process of any kind, even far from equilibrium (i.e. without adjusting $\lambda$) \cite{crooks1999entropy}. 
This perspective also allows the discovery of uncertainty principles between entropy production and any observables \cite{hasegawa2019fluctuation, horowitz2020thermodynamic}.

\subsubsection{Markov chains}

The second law can be interpreted as a rather straightforward conclusion about sequentially modeling data, which essentially states that any sequence of data and modeling distributions (modeled by a Markov chain) must increase in entropy.

A Markov chain is a time-dependent random process $m$ with values in a discrete state space $\mathcal{X}$. 
The Markov chain is parametrized by an $| \mathcal{X} | \times | \mathcal{X} |$ matrix $M$ which controls how any state distribution $P_{t} \in \Delta^{| \mathcal{X} |}$ evolves to a distribution $P_{t+1}$ at the next timestep: 
\begin{align*}
P_{t+1} = M^{\top} P_{t}
\end{align*}


When the distributions involved follow such a stochastic process -- a Markov chain over some common states -- we can deduce much more about their variation, since they are so tightly coupled by $M$. 
In fact, a straightforward calculation shows that any two distributions on the Markov chain get closer to each other as they evolve with time \cite{cover1994processes}. 
In other words, for any $i, j$, any $t$, and any distributions $P_{t}, Q_{t}$ on $\mathcal{X}$, 
\begin{align*}
\text{D} ( P_{t+1} \Vert Q_{t+1} ) \leq \text{D} ( P_{t} \Vert Q_{t} )
\end{align*}

This means that if the chain has a stationary distribution $P_{*}$, all distributions converge to it: 
$\text{D} (P_{*} \mid\mid P_{t})$ is decreasing with $t$ \cite{everett1957relative}. 

When the stationary distribution is uniform, all distributions will converge to the uniform distribution. 
Since the uniform distribution has maximal entropy, this means that entropy monotonically increases. 
This shows that entropy increases under a series of data processing operations.

We can extend this argument to reversible chains. 
A reversible chain $M$ has the following property with respect to its stationary distribution $\pi^{M}$ \cite{aldous-fill-reversible} :
\begin{align*}
\pi^{M}_{i} M_{ij} = \pi^{M}_{j} M_{ji} \qquad \forall i, j
\end{align*}
This criterion is also called detailed balance. 
Under detailed balance, free energy decreases for any distribution on the Markov chain, as shown by a straightforward calculation with the definitions (Ex. 8.12 of \cite{sethna2021statistical}). 

\subsubsection{Low-entropy initializations and the anthropic principle}

Various "Sanov-type" theorems quantify large deviation behavior, which is the probability of a sample of $n$ data points from a distribution $P$ (its empirical measure $\hat{P}_{n}$ being observed in a set $\mathcal{A}$:
$$ \frac{1}{n} \log \text{Pr} \left( \hat{P}_{n} \in \mathcal{A} \right) \leq - \min_{Q \in \mathcal{A}} \text{D} ( Q \Vert P ) $$
When in an atypical, relatively low-entropy state, this can be used on the transition measure governing evolution in time, and shows that it will typically evolve towards a more high-entropy state. 
With very high probability, there will be no return to the initial low-entropy state. 

At a basic level, this is saying that deviations towards equilibrium tend to privilege higher-entropy distributions, given some moment constraints. Other distributions will tend to have higher loss (energy). 
More atypical initial conditions will tend to degenerate to more typical ones; entropy will increase. 
Lower-entropy fluctuations will lose out to higher-entropy ones. 
If learning new features which tend to lower entropy, the system will tend to the highest-entropy of those. 
This gives a basis to the idea that informative (low-entropy) features of our world are distinctive of our particular sampled trajectory (since \cite{boltzmann1898vorlesungen}, Sec. 87) -- also called the anthropic principle \cite{weinberg1989cosmological, albert2003time}. 



\subsection{Summary of the laws}

The first and second laws are the two laws of thermodynamics referred to by \cite{khinchin1949mathematical} as quoted in the introduction. 
We have fully developed both, noting that the main content of the fundamental thermodynamic relation is in the first law. 

The third law of thermodynamics emerged by consensus later. 
In our terms, the third law is just the statement that the entropy of the zero-temperature distribution ($\lambda \to \infty$) is a finite constant. 
This is always taken to be true for the situations we consider (\cite{schrodinger1948statistical}, Ch. 3; \cite{baez2024entropy}). 
However, remember that as noted in the introduction, all of thermodynamics essentially follows from the first and second laws.  


After all this development, here is a summary of these results in the context of statistical physics. 
When modeling data with an exponential family, any change in prediction loss is caused by changing the data $f (x)$, constraints $\alpha$, or model parameters $\lambda$. 
The laws we have derived quantify how much of this change is in reducible loss (regret) versus irreducible loss ("Bayes risk" in decision theory terms). 
These are the results of thermodynamics for statistical modeling, and can all be interpreted quite rigorously in terms of their thermodynamic counterparts. 
Such analogies are a rich source of interpretation and insights, because thermodynamics is well-studied and grounded in our physical experience. 

\section{Statistical physics fluctuations in probabilistic terms}
\label{sec:fluctuations}

We have derived laws of thermodynamics that are more familiar from physics, in purely probabilistic terms. 
These track the relative impact of infinitesimal changes in the problem. 
Similar calculus tools have also been used in statistics and information geometry, to show some striking properties of exponential families which recover the laws and other physics results. 
We summarize some of these salient results, chiefly concerning fluctuations around exponential family equilibria.

\subsection{Fluctuations and heat capacity}

Differentiating $A$, with $Z := \exp (A)$: 
\begin{align*}
\frac{\partial \text{A} (\lambda)}{\partial \lambda_{i}} 
&= \frac{1}{Z} \frac{\partial Z (\lambda)}{\partial \lambda_{i}} 
= \frac{1}{Z} \mathbb{E} \left[ f_i (x) \exp \left( \sum_{i=1}^{d} \lambda_i f_i (x) \right) \right]
= \mathbb{E}_{x \sim P_{\lambda}} \left[ f_i(x) \right]
\end{align*}

For any feature, the learner observes a fixed average feature value, and considers different parameter settings $\lambda$. 
The sensitivity of the observation to parameter changes can be identified with the "heat capacity" of a particular feature:
\begin{align*}
\frac{\partial [ \mathbb{E}_{x \sim P_{\lambda}} \left[ f_i(x) \right] ]}{\partial T_i } 
= - \lambda_i^2 \frac{\partial [ \mathbb{E}_{x \sim P_{\lambda}} \left[ f_i(x) \right] ]}{\partial \lambda_i } 
= - \lambda_i^2 \text{var}_{x \sim P_{\lambda}} \left[ f_i (x) \right]
\leq 0
\end{align*}
In our language, decreasing the temperature (increasing the "coolness" $\lambda_i$) tends to raise $\mathbb{E}_{x \sim P_{\lambda}} \left[ f_i(x) \right]$. 
There is intuition for this in statistical physics, where the heat capacity is identified with $\text{var}_{x \sim P_{\lambda}} \left[ f_i (x) \right]$, going back to \cite{einstein1904allgemeinen, peliti2017einstein}.

\subsection{Quantifying fluctuations}

\subsubsection{Variations in model parameters $\lambda$}

A primary interest in AI/statistics is learning the parameters $\lambda$; how do loss-related quantities like entropy and free energy vary with $\lambda$?

One very clean aspect of this is in the free energy. 
We have already shown that $\frac{\partial \text{A} (\lambda)}{\partial \lambda_{i}} = \mathbb{E}_{x \sim P_{\lambda}} \left[ f_i(x) \right]$. 
Differentiating $\text{A}$ again, we get the Fisher information $\text{I} (\lambda)$: 
\begin{align*}
\frac{\partial^2 \text{A} (\lambda)}{\partial \lambda_{i} \partial \lambda_{j}} 
&= \frac{\partial \mathbb{E}_{x \sim P_{\lambda}} \left[ f_i (x) \right]}{\partial \lambda_{j}} 
= \frac{\partial \mathbb{E}_{x \sim P_{\lambda}} \left[ f_j (x) \right]}{\partial \lambda_{i}} \\
&= \mathbb{E}_{x \sim P_{\lambda}} \left[ f_i (x) f_j (x) \right] - \mathbb{E}_{x \sim P_{\lambda}} \left[ f_i (x) \right] \mathbb{E}_{x \sim P_{\lambda}} \left[ f_j (x) \right] \\
&= \text{cov}_{x \sim P_{\lambda}} \left[ f_i (x), f_j (x) \right] \\
&= \mathbb{E}_{x \sim P_{\lambda}} \left[ \left( f_i (x) - \mathbb{E}_{x \sim P_{\lambda}} \left[ f_i (x) \right] \right) \left( f_j (x) - \mathbb{E}_{x \sim P_{\lambda}} \left[ f_j (x) \right] \right) \right] \\
&= \mathbb{E}_{x \sim P_{\lambda}} \left[ \left( \frac{\partial [\log P_{\lambda} (x)]}{\partial \lambda_{i}} \right) \left( \frac{\partial [\log P_{\lambda} (x)]}{\partial \lambda_{j}} \right) \right] 
= - \mathbb{E}_{x \sim P_{\lambda}} \left[ \frac{\partial^2 [\log P_{\lambda} (x)]}{\partial \lambda_{i} \partial \lambda_{j}} \right] \\
&:= [\text{I} (\lambda)]_{i, j}
\end{align*}

The variation of the entropy with respect to the Lagrange multipliers depends on whether the constraints are met: 
\begin{align*}
\frac{\partial \text{H} (P_{\lambda}) }{\partial \lambda_{i} } = \mathbb{E}_{x \sim P_{\lambda}} [f_{i} (x)] - \alpha_{i}
\qquad 
\frac{\partial^{2} \text{H} (P_{\lambda}) }{\partial \lambda_{i} \partial \lambda_{j} } = \text{cov}_{x \sim P_{\lambda}} \left[ f_i (x), f_j (x) \right] - \frac{\partial \alpha_{i} }{\partial \lambda_{j} }
\end{align*}

If there is slack in the constraints, $\mathbb{E}_{x \sim P_{\lambda}} [f_{i} (x)] - \alpha_{i}$ will be nonzero, and the entropy will tend to spontaneously change. 
Otherwise, the system is at equilibrium. 
Since we know that $\text{H} (P_{\lambda})$ is convex in $\lambda$, this matrix $\frac{\partial^{2} \text{H} (P_{\lambda}) }{\partial \lambda_{i} \partial \lambda_{j} }$ is positive semidefinite.

\subsubsection{Variations in measured constraints $\alpha$}

Changes in a dataset's specified constraints $\alpha$ affect it thermodynamically. 
To investigate this, we can again look at how entropy and free energy vary with $\alpha$. 

\begin{align*}
\frac{\partial \text{H} (P_{\lambda})}{\partial \alpha_i } 
&= \sum_{j=1}^{d} \left[ \frac{\partial \text{A} (\lambda)}{\partial \lambda_{j}} \frac{\partial \lambda_{j}}{\partial \alpha_i } \right] - \frac{\partial [ \sum_{k=1}^{d} \lambda_{k} \alpha_k ]}{\partial \alpha_i } \\
&= \sum_{j=1}^{d} \left[ \frac{\partial \text{A} (\lambda)}{\partial \lambda_{j}} \frac{\partial \lambda_{j}}{\partial \alpha_i } \right] - \lambda_i - \sum_{k=1}^{d} \left[ \frac{\partial \lambda_{k}}{\partial \alpha_i } \alpha_k \right] 
= - \lambda_i
\end{align*}

This can be used as a defining property of the Lagrange multipliers, which correspond to feature-wise coolness (1/temperature) values. 
The multiplier $\lambda_{i}$ is the "shadow price" in entropy of perturbing $\alpha_{i}$. 

Differentiating this again, 
\begin{align*}
- \frac{\partial^2 \text{H} (P_{\lambda}) }{\partial \alpha_{i} \partial \alpha_{j}} 
&= \frac{\partial \lambda_{j}}{\partial \alpha_{i} } 
= \frac{\partial \lambda_{i}}{\partial \alpha_{j} } 
\end{align*}

In many situations of interest, the changes in $\alpha$ happen without changing the modeling parameters $\lambda$ at all. 
In such cases, the entropy is flat (linear) in $\alpha_{i}$, because $\frac{\partial \lambda_{j}}{\partial \alpha_{i} } = 0 = - \frac{\partial^2 \text{H} (P_{\lambda}) }{\partial \alpha_{i} \partial \alpha_{j}}$. 

The matrix $- \frac{\partial^2 \text{H} (P_{\lambda}) }{\partial \alpha_{i} \partial \alpha_{j}}$ and the Fisher information matrix $\text{I} (\lambda) = \frac{\partial^2 \text{A} (\lambda)}{\partial \lambda_{i} \partial \lambda_{j}}$ are inverses of each other, 
as can be verified by the chain rule.

The derivatives of the log-partition function with respect to the constraints $\alpha$ are as follows: 
\begin{align*}
\frac{\partial \text{A} (\lambda)}{\partial \alpha_i } 
&= \sum_{j=1}^{d} \left[ \frac{\partial \text{A} (\lambda)}{\partial \lambda_{j}} \frac{\partial \lambda_{j}}{\partial \alpha_i } \right] 
= \sum_{j=1}^{d} \left[ \frac{\partial \lambda_{j}}{\partial \alpha_i } \alpha_j \right]
\end{align*}
So the sensitivity of the free energy at equilibrium depends on $\alpha$ itself.

\subsubsection{Variations in external parameters}

Similar calculations also apply when varying an external parameter in the problem -- not $\lambda$, but some hidden variable $\gamma$ that affects $f (x)$. 
The external parameter is such that $\frac{\partial \lambda_i }{\partial \gamma} = 0$. 
Then the variation in the free energy is just a linear combination of the variation in the features: 
\begin{align*}
\frac{\partial \text{A} (\lambda)}{\partial \gamma} 
&= \frac{1}{Z} \frac{\partial Z (\lambda)}{\partial \gamma} 
= \frac{1}{Z} \mathbb{E} \left[ \left( \sum_{i=1}^{d} \frac{\partial [\lambda_i f_i (x)]}{\partial \gamma} \right) \exp \left( \sum_{i=1}^{d} \lambda_i f_i (x) \right) \right] \\
&= \sum_{i=1}^{d} \mathbb{E}_{x \sim P_{\lambda}} \left[ \frac{\partial [\lambda_i f_i (x)]}{\partial \gamma} \right]
= \sum_{i=1}^{d} \mathbb{E}_{x \sim P_{\lambda}} \left[ \lambda_i \frac{\partial [ f_i (x)]}{\partial \gamma} + \frac{\partial \lambda_i }{\partial \gamma} f_i (x) \right] \\
&= \sum_{i=1}^{d} \lambda_i \mathbb{E}_{x \sim P_{\lambda}} \left[ \frac{\partial [ f_i (x)]}{\partial \gamma} \right]
\end{align*}

A covariance-type identity allows us to calculate second-order effects of the varying parameter $\gamma$, for any pair of features $i, j$: 
\begin{align*}
\frac{\partial \mathbb{E}_{x \sim P_{\lambda}} \left[ \frac{\partial [ f_{i} (x)]}{\partial \gamma} \right]}{\partial \lambda_{j}}
= \mathbb{E}_{x \sim P_{\lambda}} \left[ \frac{\partial [ f_{i} (x)]}{\partial \gamma} \frac{\partial [ f_{j} (x)]}{\partial \gamma} \right] - \mathbb{E}_{x \sim P_{\lambda}} \left[ \frac{\partial [ f_{i} (x)]}{\partial \gamma} \right] \mathbb{E}_{x \sim P_{\lambda}} \left[ \frac{\partial [ f_{j} (x)]}{\partial \gamma} \right]
\end{align*}


\subsection{What the laws from physics offer to statistics}

Almost all elements of the laws have been quantified here. 
The work and excess heat fall out by simply differentiating the fundamental equation \eqref{eq:freeenergyexpfam}, which decomposes internal energy (loss) into free energy (regret) and entropy (Bayes loss). 
The housekeeping heat, however, does not feature in standard statistical analysis. 
Statistically, this is because it is not a true differential (as we denote with the notation $\delta Q$). 
The probabilistic laws we have developed in \Cref{sec:lawsofthermo} exactly quantify this.  

In physics conceptual terms, differentiating \eqref{eq:freeenergyexpfam} corresponds to only analyzing \emph{adiabatic} transitions. 
The laws, and especially the second, require the further development we have given.

\subsection{Equipartition of energy}

Suppose all the features $f_i (x)$ are only quadratic in the data's underlying representation $x \in \mathbb{R}^{d}$, i.e. $f_i (x)$ is a quadratic form in the coordinates of $x$. 
So rearranging the parameters $\lambda$ into a matrix $\Lambda \in \mathbb{R}^{d \times d}$, the energy describing the data $x$ can be written as a quadratic form $x^\top \Lambda x$. 
The exponential family density $P_{\Lambda}$ written accordingly: 
\begin{align*}
P_{\Lambda} (x) = \exp \left( - x^\top \Lambda x  - \text{A} (\Lambda) \right)
\end{align*}

Calculating $\text{A} (\Lambda)$, 
\begin{align*}
\text{A} (\Lambda) &= \log \int_{x \in \mathbb{R}^{d}} \exp \left( - x^\top \Lambda x \right) d x 
= \frac{1}{2} \log \left( \frac{\pi^{d}}{\det (\Lambda)} \right)
\end{align*}

Now recall that for vector parameters $\lambda$, the internal energy is 
\begin{align*}
U^{\lambda} (P_{\lambda}) = - \sum_i \lambda_i \mathbb{E}_{x \sim P_{\lambda}} [f_i (x)] = - \langle \lambda , \nabla \text{A} (\lambda) \rangle
\end{align*}

Applying this logic to the matrix $\Lambda$, in our case 
$\nabla \text{A} (\Lambda) = - \frac{1}{2} (\Lambda^{-1})^\top$, so 
\begin{align*}
U^{\Lambda} (P_{\Lambda}) = \frac{1}{2} \langle \Lambda , (\Lambda^{-1})^\top \rangle = \frac{d}{2} 
\end{align*}

This is a famous theorem from physics on the equipartition of energy: when the energy function is a quadratic form, each degree of freedom contributes a constant amount (in this case $\frac{1}{2}$; this comes down to a choice of units) to the internal energy. 
This result is an easily observed and predictable link between macroscopic and microscopic states \cite{einstein1910theorie}.

\subsection{Le Chatelier's principle}

Another deeply intuitive property of these physics problems also holds in general probabilistic terms. 
Le Chatelier's principle \cite{chatelier1884enonce} states that when part of a system is perturbed, the system changes its equilibrium to lessen the effect of the perturbation. 

This actually is a more general property of optimization problems, not specifically exponential families; the principle holds quite generally for constrained convex optimization problems \cite{leblanc1976chatelier}. 

The basic form of the principle is very simple. 
Suppose $\lambda$ and $\tau$ are two parameters in a function $\text{E} (\tau, \lambda )$ which is minimized at equilibrium. 
Denote the equilibrium $\lambda$ for a given $\tau$ as $\lambda^* (\tau) = \arg \min_{\lambda} \text{E} (\tau, \lambda )$. 
We perturb $\tau$ from $\tau_1$ to $\tau_2$, making $\lambda$ change from $\lambda_1 := \lambda^* ( \tau_1 )$ to $\lambda_2 := \lambda^* ( \tau_2 )$. 

Now by definition, $\text{E} (\tau_1 , \lambda_1 ) \leq \text{E} (\tau_1 , \lambda_2 )$. 
Similarly, $\text{E} (\tau_2 , \lambda_2 ) \leq \text{E} (\tau_2 , \lambda_1 )$. 
Subtracting one of these equations from the other, 
\begin{align*}
\text{E} (\tau_2 , \lambda_1 ) - \text{E} (\tau_1 , \lambda_1 )
\geq \text{E} (\tau_2 , \lambda_2 ) - \text{E} (\tau_1 , \lambda_2 ) 
\end{align*}

In other words, \textbf{the impact of the perturbation in $\tau$ is less under the final state $\lambda_2$ than the initial $\lambda_1$}. 
This concept is Le Chatelier's principle, and is very useful for reasoning about the second-order effects of perturbations. 
It is true here because of the convexity properties of our definitions of entropy and energy \cite{pavelka2019braun, landaulifshitz}. 
As our discussion here shows, the principle is very general, and its significance has been appreciated in a variety of scenarios in fields like economics \cite{samuelson1960extension, eichhorn1972general}.


\section{Discussion}
\label{sec:discussion}

We have seen that there is an exact quantitative correspondence between thermodynamics, as studied in physics, and 
log loss minimization problems. 
Here we further describe what the laws mean in the context of this correspondence.

\subsection{The observable setup}

What we have generally called "feature functions" are motivated in various particular ways in statistical mechanics, broadly considered "extensive" observables which acts as constraints in expectation.


In statistical mechanics, the canonical ensemble is a powerful tool for making predictions about the behavior of systems \cite{landaulifshitz}. 
In our setup, it amounts to privileging one feature observation -- energy -- and taking the corresponding exponential family, with some fixed parameters $\lambda$ \cite{pathria2017statistical}.
The maximum entropy principle is a generalization of the canonical ensemble to arbitrary constraints, on the basis that the entropy-maximizing distribution subject to a set of constraints is the one that is most consistent with the constraints and the least biased. 
Our laws for exponential families implicitly use this principle. 

The exponential family can be motivated as the most likely distribution to have generated the given observations (\Cref{sec:basicexpfam}). 
This was the original spirit of Boltzmann's development, and this is the reason the laws are derived around an exponential family modeling equilibrium $P_{\lambda}$. 
Such an equilibrium is uniquely motivated from first principles.

\subsection{Reversible shifts}

In the general case, a distribution shift can change an entire dataset -- all the feature representations and labels of every single example. 
Each data point must be taken separately to exponentially increase the size of the search space, because each can be changed individually apart from the others. 

The act of observation, of average values of feature functions, is how we process information about the world. 
We have shown that when we observe $d$ feature functions, the learning problem is effectively $d$-dimensional, as we are optimizing over the corresponding exponential family. 
For some fixed features, the exponential family traces out a $d$-dimensional sub-manifold (often called the Gibbs manifold) of the possible models of the underlying dataset. 

Rather than modeling an arbitrary change in the data distribution, which can scale very badly in the sample size $n$, tracking changes over this lower-dimensional sub-manifold is much more tractable.  
We can think about exactly retracing such a change by infinitesimal changes in the parameters $\lambda$. 
Following thermodynamics, such shifts are called \textbf{reversible}. 

A reversible change in distributions progresses within the manifold of equilibrium states, since these states are so simply characterized by a few variables $\lambda$. 
Any such shift can be reversed by controlling those variables. 
It is a thermodynamic process carried out arbitrarily slowly -- enough so that the system is in equilibrium at every instant. 

Natural shifts in general, on the other hand, move out of that manifold through much more complex non-equilibrium states. 
In general, this requires an enormous number of variables (perhaps the entire dataset) to characterize these states.

\subsection{Heat, work, and thermodynamic cycles}

The general probabilistic formulation we have taken makes it very clear that any change in the data or constraints will decompose into only work and heat terms. 
The predicted distribution of the data is determined completely by the data's energy levels, so work -- the change in those -- is meaningfully easier to adapt to, and free energy evidently represents (up to a constant) the amount of energy usable for work. 
The difficulty is to account for changes in constraints that are unaccounted for by the existing data. 
This is what heat represents in this context -- the change in performance (internal energy) that is not due to the change in the data's energy levels (predictions). 


The housekeeping heat \cite{speck2005integral} is associated with maintaining the non-equilibrium state in which $\mathbb{E}_{x \sim P_{\lambda}} [f (x)] \neq \alpha$, even without changing the model $\lambda$. 
The housekeeping heat is zero for \textbf{adiabatic} shifts, defined by $\delta Q_i^{\lambda} = 0 \;\;\forall i$. 
For adiabatic shifts, the change in the data $\Delta (x)$ is the only source of the change in the constraints $\Delta^{\alpha}$. 

Meanwhile, the excess heat \cite{seifert2005entropy, seifert2012stochastic} is associated with changing $\lambda$ in a non-equilibrium state. 
So it is zero for reversible shifts.

On a related note, a notion of specific heat of a feature $i$ in a dataset is: 

$$ T_{i} \frac{\partial \text{H} (P_{\lambda})}{\partial T_{i}} = - \lambda_{i} \frac{\partial \text{H} (P_{\lambda})}{\partial \lambda_i} = - \lambda_{i} \left( \mathbb{E}_{x \sim P_{\lambda}} [f_{i} (x)] - \alpha_{i} \right) $$



Thermodynamic cycles were studied during the design and development of engines in the 19th century. 
These are comprised of the same few types of thermodynamic shifts, which are useful in modeling scenarios also: adiabatic shifts in which the housekeeping heat is zero, and other shifts in which entropy, parameters $\lambda$, or features $f_{i}$ are held constant (respectively isentropic, isothermal, isochoric). 

Thermodynamic cycles may be interesting to statistical modelers for the same reason they are interesting to physicists: we want to extract work from energy with minimum heat. 
In our case, this refers to our desire to change reducible loss rather than irreducible loss, when changing the modeling problem's total loss (internal energy). 
Similarly, we also find traditional thermodynamic motivations relevant in modeling the processes in a cycle as quasistatic; this reduces the problem to a tractable sub-manifold of models of the dataset. 

In physics, an idealized engine is considered a thermodynamic cycle in which changes of energy are decomposed into heat and work components. 
Efficiencies in avoiding waste heat, in cycles like the Carnot and Otto cycles, are of likely interest in modeling scenarios as well.

There is an thermodynamic cost to memory-limited computation, that prevents it from being fully reversible and "frictionless." This intimately bridges the information-theoretic and thermodynamic notions of entropy, and has been explored through various perspectives \cite{bennett1982thermodynamics, zurek1989algorithmic, szilard1964decrease, landauer1961irreversibility}. 
The implications of the general probabilistic laws of this paper on such calculations would vary depending on the dataset, an aspect which can be explored in future work.

\newpage
\bibliographystyle{alpha}
\bibliography{main}

\newpage
\appendix

\section{Some basic properties of exponential families}
\label{sec:basicexpfam}

Here we collect some properties of exponential families that are salient in light of the laws.

\subsection{Mean-field approximation: Bogoliubov's inequality}

Exponential families have another convenient property for learning. 
Suppose we are looking to approximate an exponential family distribution with another "\textit{variational}" model distribution, possibly in a different exponential family, with different features and parameters $\psi$. 

As part of an exponential family, we can write this $P_{\psi} (x) = \exp \left( - U^{\psi} (x) - A (\psi) \right)$. 
For any such variational $P_{\psi}$, we have 
\begin{align*}
0 &\leq \text{D} (P_{\psi} \Vert P_{\lambda}) \\
&= \mathbb{E}_{x \sim P_{\psi} } \left[ - U^{\psi} (x) - A (\psi) + U^{\lambda} (x) + A (\lambda) \right] \\
&= (U^{\lambda} (P_{\psi}) - U^{\psi} (P_{\psi})) + (F^{\psi} (P_{\psi}) - F^{\lambda} (P_{\lambda}))
\end{align*}

Thus, if $\psi$ is chosen to have $U^{\lambda} (P_{\psi}) = U^{\psi} (P_{\psi})$, then the free energy $F^{\psi} (P_{\psi}) \geq F^{\lambda} (P_{\lambda})$. 
In this case, the free energy of $P_{\psi}$ can be used as a tight bound on the free energy of the unknown distribution $P_{\lambda}$, as long as $U^{\lambda}$ and $U^{\psi}$ are the same under the variational distribution $P_{\psi}$. 

Similarly, a variational lower bound $F^{\psi} (P_{\psi}) \leq F^{\lambda} (P_{\lambda})$ holds as long as $U^{\lambda}$ and $U^{\psi}$ are the same under the target distribution $P_{\lambda}$. 
\begin{align*}
0 &\geq - \text{D} (P_{\lambda} \Vert P_{\psi}) \\
&= - \mathbb{E}_{x \sim P_{\lambda} } \left[ - U^{\lambda} (x) - A (\lambda) + U^{\psi} (x) + A (\psi) \right] \\
&= (U^{\lambda} (P_{\lambda}) - U^{\psi} (P_{\lambda})) + (F^{\psi} (P_{\psi}) - F^{\lambda} (P_{\lambda}))
\end{align*}

Note that conditions on $U (P)$ are often more amenable to computation than $H (P)$ or $F (P)$, since expectation values of observables can be typically computed easily from finite samples. 

This is a basic principle from statistical mechanics underlying mean-field variational inference. 
It is often called Bogoliubov's inequality.





\subsection{Data-generating "robustness"}

Exponential families are remarkably easy to compare to each other with the divergence $\text{D} ( \cdot \Vert \cdot )$.  
For any distributions $P_{\alpha}, P_{\beta}$ from the same exponential family $\mathcal{Q}$, and any distribution $Q$, we have 

\begin{align*}
\text{D} (Q \Vert P_{\beta}) - \text{D} (Q \Vert P_{\alpha}) 
&= \mathbb{E}_{x \sim Q} \left[ \log \left( P_{\alpha} (x) \right) - \log \left( P_{\beta} (x) \right) \right] \\
&= \mathbb{E}_{x \sim Q} \left[ \left( \sum_{i=1}^{d} \alpha_i f_i(x) - A (\alpha) \right) - \left( \sum_{i=1}^{d} \beta_i f_i(x) - A (\beta) \right) \right] \\
&= \sum_{i=1}^{d} (\alpha_i - \beta_i ) \mathbb{E}_{x \sim Q} \left[ f_i(x) \right] - A (\alpha) + A (\beta)
\end{align*}

This key equation has a few important consequences when $Q \in \mathcal{A}$. 
Suppose $Q, P_{\alpha} \in \mathcal{A}$, i.e. the data follow the same moment constraints as one of the distributions. 
Then, 
\begin{align}
\label{eq:expfamrobustness}
\text{D} (Q \Vert P_{\beta}) - \text{D} (Q \Vert P_{\alpha}) 
&= \sum_{i=1}^{d} (\alpha_i - \beta_i ) \mathbb{E}_{x \sim Q} \left[ f_i(x) \right] - A (\alpha) + A (\beta) \nonumber \\
&= \sum_{i=1}^{d} (\alpha_i - \beta_i ) \mathbb{E}_{x \sim P_{\alpha}} \left[ f_i(x) \right] - A (\alpha) + A (\beta) \nonumber \\
&= \mathbb{E}_{x \sim P_{\alpha}} \left[ \log \left( \frac{P_{\alpha} (x)}{P_{\beta} (x)} \right) \right] 
= \text{D} (P_{\alpha} \Vert P_{\beta})
\end{align}

This is called "robustness" of exponential families \cite{grunwald2007minimum}: 
the relative performance of two coding schemes $P_{\alpha}, P_{\beta}$ is the same when measured by any $Q \in \mathcal{A}$. 
In our situation, it means that if $\hat{P}_{n}$ denotes the observed data distribution and $P_{\mathcal{A}}^{*}$ the max-entropy distribution under the observed features, 
then for any $P_{\lambda} \in \mathcal{Q}$, 
\begin{align*}
\text{D} (\hat{P}_{n} \Vert P_{\lambda}) - \text{D} (\hat{P}_{n} \Vert P_{\mathcal{A}}^{*}) 
&= \text{D} (P_{\mathcal{A}}^{*} \Vert P_{\lambda})
\end{align*}

In other words, for the task of predicting the observed $\hat{P}_{n}$, the relative performance of any model $P_{\lambda}$ to the best $P_{\mathcal{A}}^{*}$ is just $\text{D} (P_{\mathcal{A}}^{*} \Vert P_{\lambda})$. 
The relative prediction loss of any exponential family distribution $P_{\lambda}$ to the best $P_{\mathcal{A}}^{*}$ is $\text{D} (P_{\mathcal{A}}^{*} \Vert P_{\lambda})$, regardless of any other details of the data $\hat{P}_{n}$. 

This has been shown with "performance" being measured by regret. 
Note that  
\begin{align*}
\text{D} (Q \Vert P_{\beta}) - \text{D} (Q \Vert P_{\alpha})
= 
\text{H} (Q , P_{\beta}) - \text{H} (Q , P_{\alpha})
\end{align*}
so all these statements are true for relative loss as well.

\subsection{Approximation and estimation error}
\label{sec:subappest}



Setting $P_{\alpha}$ to be $P_{\mathcal{A}}^{*} \in \mathcal{A} \cap \mathcal{Q}$ in the equation \eqref{eq:expfamrobustness} above, we get a very useful result about this max-entropy distribution $P_{\mathcal{A}}^{*}$: the divergence satisfies a Pythagorean theorem for any $P$ meeting the moment constraints ($P \in \mathcal{A}$), and any $P_{\lambda} \in \mathcal{Q}$. 
\begin{align*}
\forall P \in \mathcal{A} , P_{\lambda} \in \mathcal{Q} : \qquad
\underbrace{\text{D} (P \Vert P_{\lambda}) }_{\text{regret}} = \underbrace{\text{D} (P_{\mathcal{A}}^{*} \Vert P_{\lambda})}_{\text{estimation error}} + \underbrace{\text{D} (P \Vert P_{\mathcal{A}}^{*})}_{\text{approximation error}}
\end{align*}

This is a decomposition of the relative loss (the regret) into estimation and approximation errors. 

\begin{itemize}
\item
The approximation error is lowered by considering more expressive architectures. 
\item
The estimation error is lowered by considering more data. 
\end{itemize}

This means that if all we know about the data is encapsulated in $\mathcal{A}$, it is a good idea to minimize over the parametric family $\mathcal{Q}$ (under the geometry induced by $\text{D}$). 
There are some specific consequences to the Pythagorean equality above. 

In the Pythagorean equality, we clearly see that both the approximation and estimation errors are $\geq 0$. 
Applying this understanding gives us two inequalities, which hold for any $\mathcal{A}$ and associated $\lambda$. 

First, the overall regret exceeds the estimation error: 
\begin{align*}
\text{D} (P \Vert P_{\lambda}) \geq \text{D} (P_{\mathcal{A}}^{*} \Vert P_{\lambda}) \quad \forall P \in \mathcal{A} , P_{\lambda} \in \mathcal{Q}
\end{align*}

This can be readily interpreted -- 
for any exponential family model $P_{\lambda}$, the actual data is harder to encode than the max-ent distribution. 

On the other hand, the overall regret also evidently exceeds the approximation error: 
\begin{align*}
\text{D} (P \Vert P_{\lambda}) \geq \text{D} (P \Vert P_{\mathcal{A}}^{*}) \quad \forall P \in \mathcal{A} , P_{\lambda} \in \mathcal{Q}
\end{align*}

Since the data $\hat{P}_{n} \in \mathcal{A}$ by definition, this applies to them: 
$\text{D} (\hat{P}_{n} \Vert P_{\lambda}) \geq \text{D} (\hat{P}_{n} \Vert P_{\mathcal{A}}^{*}) \quad \forall \lambda$. 
As $P_{\mathcal{A}}^{*}$ is in the exponential family $\mathcal{Q}$, this means that 
\begin{align*}
P_{\mathcal{A}}^{*} = \arg\min_{P_{\lambda} \in \mathcal{Q}} \; \text{D} (\hat{P}_{n} \Vert P_{\lambda})
\end{align*}
which shows that $P_{\mathcal{A}}^{*}$ minimizes the log loss (cross entropy) to the data over $\mathcal{Q}$.

\subsection{The estimation error and deviance}

The estimation error relates to the concept of \textbf{deviance}, which uses the divergence $\text{D} (\cdot \Vert \cdot)$ to relate a non-equilibrium probability distribution $P_{\lambda}$ to the equilibrium distribution $P_{\lambda^{*}}$. 
We can evaluate the ratio of the distributions at any given observed feature representation $f(x) \in \mathbb{R}^{d}$: 
\begin{align*}
\log \left( \frac{P_{\lambda} (f(x))}{P_{\lambda^{*}} (f(x))} \right)
= \sum_{i=1}^{d} (\lambda_i - \lambda_i^* ) f_i(x) + A (\lambda^{*}) - A (\lambda)
\end{align*}
At the actual observation $f(x) = \alpha$, 
\begin{align*}
\log \left( \frac{P_{\lambda} (\alpha)}{P_{\lambda^{*}} (\alpha)} \right) 
&= \sum_{i=1}^{d} (\lambda_i - \lambda_i^* ) \alpha_i + A (\lambda^{*}) - A (\lambda) 
= - \text{D} (P_{\lambda^{*}} \Vert P_{\lambda}) 
\end{align*}
which exactly shows how suboptimal parameter settings will deviate around the optimum in modeling the observations. 

Solving for the density $P_{\lambda}$, we get "Hoeffding's formula" for the density at the observations $\alpha$:
\begin{align*}
P_{\lambda} (\alpha)
&= P_{\lambda^{*}} (\alpha) \exp \left( - \text{D} (P_{\lambda^{*}} \Vert P_{\lambda}) \right)
\end{align*}

\subsection{The approximation error and entropy}
\label{sec:subapproxerror}

How well does the information projection $P_{\mathcal{A}}^{*}$ approximate the data $P$?

For any data meeting the constraints, i.e. $P \in \mathcal{A}$, and any $\lambda$: 
\begin{align*}
\text{D} (P \Vert P_{\mathcal{A}}^{*}) 
&= \text{D} (P \Vert P_{\lambda}) - \text{D} (P_{\mathcal{A}}^{*} \Vert P_{\lambda}) 
= \text{H} (P, P_{\lambda}) - \text{H} (P_{\mathcal{A}}^{*} , P_{\lambda}) + \text{H} (P_{\mathcal{A}}^{*}) - \text{H} (P) 
\end{align*}

In particular this is true for $\lambda = 0$, in which case $P_{\lambda} = P_{\lambda = 0} = P_{0}$ (the uniform distribution over the data), 
so $\text{H} (P, P_{0}) = \text{H} (P_{\mathcal{A}}^{*} , P_{0})$, and this reduces to 
$$
\text{D} (P \Vert P_{\mathcal{A}}^{*}) 
= \text{H} (P_{\mathcal{A}}^{*}) - \text{H} (P) 
$$ 

Therefore, the data will be well approximated if they have roughly maximal entropy under the constraints. 
Tying together these concepts, the max-entropy problem has a variational characterization: $\text{H} (P_{\lambda^{*}}) - \text{H} (P) = \text{D} (P \Vert P_{\lambda^{*}})$ for all $P$ matching the moment constraints. 
This extends to any moment constraints, so we could also say for any $\lambda$ that 
$
\text{D} (P \Vert P_{\lambda}) = \text{H} (P_{\lambda}) - \text{H} (P)
$, 
for all $P$ having the same feature moments as $P_{\lambda}$.

\subsection{Evaluating exponential family models}

Using this in the regret decomposition above, 
\begin{align*}
\underbrace{\text{D} (P \Vert P_{\lambda}) }_{\text{regret}} &= \text{D} (P_{\mathcal{A}}^{*} \Vert P_{\lambda}) + \text{H} (P_{\mathcal{A}}^{*}) - \text{H} (P) \\
&= \text{H} (P_{\mathcal{A}}^{*} , P_{\lambda}) - \text{H} (P) 
\end{align*}

Adding $\text{H} (P)$ to both sides gives an interesting result: 
\begin{align*}
\forall P \in \mathcal{A} : \qquad \qquad \text{H} (P , P_{\lambda}) &= \text{H} (P_{\mathcal{A}}^{*} , P_{\lambda})
\end{align*}

The interpretation here is unambiguous: for evaluating the loss using the exponential family $\mathcal{Q}$, we can pretend the data follows $P_{\mathcal{A}}^{*}$.

\subsection{Using data to approximate the exponential family}

We can flip the roles of $P$ and $P_{\mathcal{A}}^{*}$ in the above question about divergence: how well does the data $P$ approximate $P_{\mathcal{A}}^{*}$?

It turns out that: 
\begin{align*}
- \text{D} (P_{\mathcal{A}}^{*} \Vert P) 
\geq \frac{1}{n} \log \text{Pr} ( \hat{P}_{n} \in \mathcal{A} ) 
\geq - \text{H} (P_{\mathcal{A}}^{*} , P )
\end{align*}

So if $\hat{P}_{n}$ is consistent with the observations $\mathcal{A}$ and $\text{Pr} ( \hat{P}_{n} \in \mathcal{A} )$ is high, then $\text{D} (P_{\mathcal{A}}^{*} \Vert P)$ is quite low - the data $P$ is a good approximation of samples generated with $P_{\mathcal{A}}^{*}$. 

To show the lower bound here, we use a Sanov-type probability identity \cite{balsubramani2020sharp} and the fact that $\mu_{\mathcal{A}} \in \mathcal{A}$: 
\begin{align*}
\frac{1}{n} \log \text{Pr} ( \hat{P}_{n} \in \mathcal{A} ) 
&= - \text{D} (P_{\mathcal{A}}^{*} \Vert P ) - \frac{1}{n} \text{D} ( \mu_{\mathcal{A}} \Vert P_{\mathcal{A}}^{*n} ) \\
&= - \text{H} (P_{\mathcal{A}}^{*} , P ) + \text{H} ( P_{\mathcal{A}}^{*} ) - \frac{1}{n} \text{H} ( \mu_{\mathcal{A}} , P_{\mathcal{A}}^{*n} ) + \frac{1}{n} \text{H} ( \mu_{\mathcal{A}} ) \\
&= - \text{H} (P_{\mathcal{A}}^{*} , P ) + \frac{1}{n} \text{H} ( P_{\mathcal{A}}^{*n} ) - \frac{1}{n} \text{H} ( P_{\mathcal{A}}^{*n} ) + \frac{1}{n} \text{H} ( \mu_{\mathcal{A}} ) 
\geq - \text{H} (P_{\mathcal{A}}^{*} , P )
\end{align*}


\subsection{The log-partition function and higher moments}

A well-known result \cite{chowdhury2023bregman} connects the cumulant-generating function of the features under an exponential family distribution $P_{\lambda} \propto \exp \left( \sum_{i=1}^{d} \lambda_i f_i(x) \right)$ to the log-partition function $A (\lambda)$. 
\begin{align*}
\log \mathbb{E}_{x \sim P_{\lambda}} \left[ \exp \left( \sum_{i=1}^{d} \theta_i f_i(x) \right) \right] 
&= A (\lambda + \theta) - A (\lambda)
\end{align*}
This also implies a cumulant-generating function for the centered features, i.e. those with mean zero, which is in the form of a Bregman divergence $\text{B}_{F} (P, Q) := F (P) - F (Q) - (P-Q)^\top \nabla F (Q)$: 
\begin{align*}
\log \mathbb{E}_{x \sim P_{\lambda}} \left[ \exp \left( \sum_{i=1}^{d} \theta_i \left( f_i(x) - \mathbb{E}_{x \sim P_{\lambda}} [f_i (x)] \right) \right) \right] 
&= A (\lambda + \theta) - A (\lambda) - \theta^\top \nabla A (\lambda)
= \text{B}_{A} (\lambda + \theta, \lambda)
\end{align*}




\end{document}